\documentclass[onecolumn,11pt,draftcls]{IEEEtran}
\usepackage{amsmath,amssymb,amsthm,graphicx,graphics,subfigure,epsfig,enumerate}
\usepackage{parskip}		
\usepackage{bm}
\usepackage{color}

\newcommand{\pp}[1]{{\color{black}{#1}}}

\title{Modal Analysis Using Sparse and Co-prime Arrays}
\author{Pooria Pakrooh,~\IEEEmembership{Student Member,~IEEE,}  Louis L. Scharf,~\IEEEmembership{Life Fellow,~IEEE}, \pp{and} Ali~Pezeshki,~\IEEEmembership{Member,~IEEE} \vfill\vfill
\thanks{This work is supported in part by NSF under grants CCF-1018472 and CCF-1422658.} \thanks{A preliminary version of a subset of the results reported here \pp{was} presented at the 2014 Asilomar Conference on Signals, Systems, and Computers, Pacific Grove, CA, Nov. 2-5, 2014, in a paper entitled ``Characterization of orthogonal subspaces for alias-free reconstruction of damped complex exponential modes in sparse arrays,'' by P. Pakrooh, A. Pezeshki, and L. L. Scharf.}
\thanks{P. Pakrooh is with the Department of Electrical and
Computer Engineering, Colorado State University, Fort Collins, CO 80523, USA (e-mail: Pooria.Pakrooh@colostate.edu).}
\thanks{L. L. Scharf is with the Department of Mathematics and the Department of Statistics, Colorado State University, Fort Collins, CO 80523, USA (e-mail: Louis.Scharf@colostate.edu).}
\thanks{A. Pezeshki is with the Department of Electrical and
Computer Engineering and the Department of Mathematics, Colorado State University, Fort Collins, CO 80523, USA (e-mail: Ali.Pezeshki@colostate.edu).}

\thanks{}
\thanks{}
\thanks{}
\thanks{}}\newcommand\remark[1]{\textit{Remark #1:}}
\newcommand{\yb}{{\bf y}}
\newcommand{\Ab}{{\bf A}}
\newcommand{\Bb}{{\bf B}}
\newcommand{\Cb}{{\bf C}}

\newcommand{\eb}{{\bf e}}
\newcommand{\xb}{{\bf x}}

\newcommand{\Vb}{{\bf V}}
\newcommand{\zb}{{\bf z}}

\newcommand{\Gb}{{\bf G}}

\newcommand{\Yb}{{\bf Y}}

\begin{document}

\maketitle

\begin{abstract}
Let a measurement consist of a linear combination of damped complex exponential modes, plus noise. The problem is to estimate the parameters of these modes, as in line spectrum estimation, vibration analysis, speech processing, system identification, and direction of arrival estimation. Our results differ from standard results of modal analysis to the extent that we consider sparse and co-prime samplings in space, or equivalently sparse and co-prime samplings in time. Our main result is a characterization of the orthogonal subspace. This is the subspace that is orthogonal to the signal subspace spanned by the columns of the generalized Vandermonde matrix of modes in sparse or co-prime arrays. This characterization is derived in a form that allows us to adapt modern methods of linear prediction and approximate least squares, such as iterative quadratic maximum likelihood (IQML), for estimating mode parameters. Several numerical examples are presented to demonstrate the validity of the proposed modal estimation methods, and to compare the fidelity of modal estimation with sparse and co-prime arrays, versus SNR. Our calculations of Cram\'{e}r-Rao bounds allow us to analyze the loss in performance sustained by sparse and co-prime arrays that are compressions of uniform linear arrays.


\end{abstract}

\begin{keywords}
Co-pime array, IQML, modal analysis, orthogonal subspaces, sparse array \vspace{-.2cm}
\end{keywords}

\section{Introduction}

In this paper, we investigate the problem of estimating the parameters of damped complex exponentials from the observation of non-uniform samples of their weighted sum. This problem arises in many applications such as modal analysis, speech processing, system identification, and direction of arrival (DOA) estimation.

There is a vast literature on different modal estimation methods from uniformly sampled time or space series data, starting with the work of Prony \cite{Prony1795}. Other methods include approximate least squares or maximum likelihood estimation \cite{Kumaresan86}, \cite{Bresler86}, reduced rank linear prediction \cite{Kumaresan83}, \cite{Kumaresan84}, MUSIC \cite{Schmidt86}, and ESPRIT \cite{paulraj85}, \cite{Roy89}. While there are extensions of MUSIC and ESPRIT for direction of arrival estimation from non-uniformly sampled data (see, e.g., \cite{Belloni2006}-\nocite{Gershman2009}\nocite{Friedlander1992}\nocite{Friedlander1993}\cite{Gershman2005}), Prony-like methods have mainly been developed for uniformly sampled data, and extending such methods to non-uniformly sampled data has not received much attention (exceptions being \cite{coluccio2007} and \cite{Peter2011}).

Non-uniform sensor array geometries, without aliasing ambiguities, have a long history in sensor array processing, dating back to minimum-redundancy arrays \cite{Moffet68}. The introduction of co-prime arrays in \cite{coprime1}, \cite{coprime2}, and \cite{coprime3}  has created renewed interest in such geometries. In this paper, we consider two specific cases of non-uniform sensor arrays. These are sparse arrays and co-prime arrays. Both of these geometries can be viewed as subsampled (or compressed) versions of a dense uniform line array, whose consecutive elements are separated by a half wavelength in space.\footnote{If we were sampling in time, then the dense sequence of uniform samples would have had spacings equal to the Nyquist interval.} Specifically, the sparse array can be thought of as a subsampled version of a dense uniform line array, plus an extra sensor that is positioned at a location on the array that allows us to resolve aliasing ambiguities. The co-prime array consists of two uniform subarrays, each obtained by uniformly subsampling a dense uniform line array with co-prime subsampling factors. The co-prime property allows for resolving aliasing ambiguities.

Naturally, any subsampling in space results in a reduction in signal-to-noise ratio (SNR), by the compression factor, and leads to a loss in estimation performance. Our studies in \cite{Pakrooh13a}, \cite{Pakrooh-FITSP}, and \cite{Pakrooh13b} address the effect of compression on Fisher information, the Cram\'{e}r-Rao bound, and the probability of a swap between signal and noise subspaces. Assuming that the loss in SNR due to compression has tolerable effects on estimation or detection, or can be compensated by collecting more temporal snapshots (requiring a scene to remain stationary for a longer period), the question is how can methods of linear prediction and approximate least squares be adapted to the estimation of mode parameters in sparse and co-prime arrays? In this paper, we address this question.

We determine a parameterization of the orthogonal subspace. This is the subspace  that is orthogonal to the signal subspace spanned by the columns of a generalized Vandermonde matrix of the modes in sparse and co-prime arrays. This parameterization is of a form that is particularly suitable for utilizing approximate least squares, such as iterative quadratic maximum likelihood (IQML) (see \cite{Kumaresan86}, \cite{Bresler86}, and \cite{McClellan91}), for estimating the modes. Although we present our numerical results in the context of sensor array processing, all of our results apply to the estimation of complex exponential modes from time series data. Our numerical results here, and in \cite{Pakrooh13a}, \cite{Pakrooh-FITSP}, and \cite{Pakrooh13b} , show that there is a loss in performance sustained by sparse and co-prime arrays that are compressions of uniform linear arrays. A rough rule of thumb is that effective SNR is reduced by $10\mathrm{log}_{10}C$, where $C$ is the compression ratio. For example, in our experiments a $50$-element array is subsampled to a $14$-element co-prime array, for a compression ratio of  $50/14$. The loss in SNR is roughly $5.5$ dB.

\remark{1} A small number of other authors have also considered estimating the parameters of complex exponentials from non-uniformly sampled data using Prony-like methods. In \cite{coluccio2007}, the authors approach the modal estimation problem by fitting a polynomial to the non-uniform samples and estimating the parameters of the exponentials using linear regression. For the case that the modes are on the unit circle, in \cite{Peter2011} a truncated window function is fitted to the non-uniform measurements in the least squares sense, and then an approximate Prony method is proposed to estimate the frequencies of the exponentials. These approaches are different from ours and do not involve characterization of orthogonal subspaces for utilizing modern methods of linear prediction.

\section{Problem Statement}\label{sec:PS}

Consider a non-uniform line array of $m$ sensors at locations $\mathbb{I}=\{i_0,i_1,\ldots, i_{m-1}\}$ in units of half wavelength in space. We assume, without loss of generality, that $i_0=0$. Suppose the array is observing a weighted superposition of $p$ damped complex exponentials (modes). These modes are determined by the mode parameters $z_k=\rho_k e^{j\theta_k}$, $k=1,2, \ldots, p$, where the $k$th mode has a damping factor $\rho_k$ and an electrical angle $\theta_k\in (-\pi,\pi]$.  Suppose the array collects $N$ temporal snapshots. Then, the measurement equation for the $l$th sensor (located at $i_l$) can be written as
\begin{equation}
y_l[n]=\sum_{k=1}^{p}x_k[n]z_k^{i_l}+e_l[n], \quad n=0,1,\ldots, N-1,
\end{equation}
where $n$ is the snapshot index, $x_k[n]$ denotes the amplitude (or weight) of the $k$th mode at index $n$, and $e_l[n]$ is the measurement noise at sensor $l$. In vector form, we have $\yb[n]\in\mathbb{C}^m$,
\begin{equation}\label{eq:Main1}
\yb[n]=\Vb(\zb,\mathbb{I})\xb[n]+\eb[n], \quad n=0,1,\ldots, {N-1},
\end{equation}
where $\yb[n]=[y_0[n],y_1[n],\ldots,y_{m-1}[n]]^T$ is the array measurement vector, $\xb[n]=[x_1[n],x_2[n],\break\ldots,x_{p}[n]]^T$ is the vector of mode amplitudes at index $n$, $\eb[n]=[e_{0}[n],e_{1}[n],\ldots,e_{m-1}[n]]^T$ is the noise vector at index $n$, and $\Vb(\zb,\mathbb{I})\in\mathbb{C}^{m\times p}$ is a generalized Vandermonde matrix \pp{of the modes $\zb=[z_1,z_2,\dots,z_p]^T$}, given by
\begin{equation}\label{eq:Main2}
\Vb(\zb,\mathbb{I})=\begin{bmatrix}
  z_1^{i_0} & z_2^{i_0} & \cdots & z_p^{i_0} \\
  z_1^{i_1} & z_2^{i_1} & \cdots & z_p^{i_1} \\
  \vdots  & \vdots  & \ddots & \vdots  \\
  z_1^{i_{m-1}} & z_2^{i_{m-1}} & \cdots & z_p^{i_{m-1}}
 \end{bmatrix}.
\end{equation}

\begin{figure}[ht]\centering
\includegraphics[width=168pt]{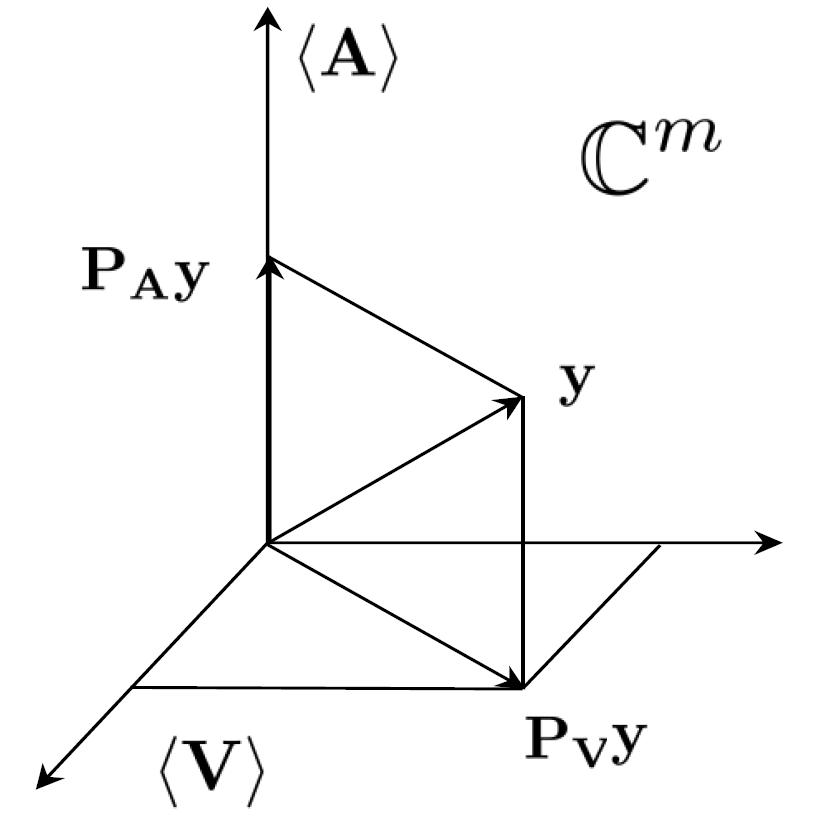}
\center
  \caption{The signal subspace $\langle \Vb(\zb,\mathbb{I}) \rangle$ and the orthogonal subspace $ \langle \Ab(\zb,\mathbb{I}) \rangle=\langle \Vb(\zb,\mathbb{I}) \rangle^\perp$. In the figure, we have dropped $(\zb, ,\mathbb{I})$ and have simply used $\Ab$, $\Vb$, $\langle \Ab \rangle$, and $\langle \Vb \rangle$.}\label{fig:Subspaces}
\end{figure}

We consider the case where $\xb[n]$ is free to change with $n$, and assume that the $\eb_l[n]$'s, are i.i.d. complex normal with mean zero and \pp{variance $\sigma^2$}. This means that the measurement vectors $\yb[n]$, $n=0,1,\ldots,N-1$ are i.i.d proper complex normal with mean $\Vb(\zb,\mathbb{I})\xb[n]$ and covariance $\sigma^2\mathbf{I}$. Under this measurement model, the least squares estimation and the maximum likelihood estimation of the modes $\{z_k\}_{k=1}^p$ and mode weights  $\{\xb[n]\}_{n=0}^{N-1}$ are equivalent and can be posed as
\begin{equation}
\min_{\zb,\xb[0],\dots,\xb[N-1]}\sum_{n=0}^{N-1}\|\yb[n]-\Vb(\zb,\mathbb{I})\xb[n]\|_2^2.
\end{equation}
The least squares estimate of $\xb[n]$ is
\begin{equation}\label{eq:LS_W}
\hat{\xb}[n]= {\Vb^+(\zb,\mathbb{I})}\yb[n],
\end{equation}
where $\Vb^+ (\zb,\mathbb{I})= (\Vb^H(\zb,\mathbb{I})\Vb(\zb,\mathbb{I}))^{-1}\Vb^H(\zb,\mathbb{I})$ is the Moore-Penrose pseudoinverse of $\Vb(\zb,\mathbb{I})$. The least squares estimate of the modes is obtained as
\begin{align}\label{eq:min}
\hat{\zb}&=\arg\min_{\zb}\sum_{n=0}^{N-1}\yb^H[n](\mathbf{I}-\mathbf{P}_{\Vb(\zb,\mathbb{I})})\yb[n]\nonumber\\ &=\arg\min_{\zb}\sum_{n=0}^{N-1}\yb^H[n]\mathbf{P}_{\Ab(\zb,\mathbb{I})}\yb[n],
\end{align}
where $\Ab(\zb,\mathbb{I})$ is a full column rank matrix that satisfies
\begin{equation}
\Ab^H(\zb,\mathbb{I})\Vb(\zb,\mathbb{I})=\mathbf{0}_{(m-p)\times p},
\end{equation}
and $\mathbf{P}_{\Vb(\zb,\mathbb{I})}$ and $\mathbf{P}_{\Ab(\zb,\mathbb{I})}=\mathbf{I}-\mathbf{P}_{\Vb(\zb,\mathbb{I})}$ are the orthogonal projections onto the column spans of $\Vb(\zb,\mathbb{I})$ and $\Ab(\zb,\mathbb{I})$, respectively. We denote these column spans by the subspaces $\langle \Vb(\zb,\mathbb{I}) \rangle$ and $\langle \Ab(\zb,\mathbb{I}) \rangle$. We call $\langle \Vb(\zb,\mathbb{I}) \rangle$ the {\em signal subspace} and $\langle \Ab(\zb,\mathbb{I}) \rangle$ the {\em orthogonal subspace}. Note that $\langle \Ab(\zb,\mathbb{I}) \rangle=\langle \Vb(\zb,\mathbb{I}) \rangle^\perp$. See Figure \ref{fig:Subspaces}.

For a given array geometry, the {\em basis matrix}
$\Vb(\zb,\mathbb{I})$ given in (\ref{eq:Main2}), and the subspace $\langle\Vb(\zb,\mathbb{I})\rangle$, are fully characterized by the $p$ modes $\zb=[z_1,z_2,\dots,z_p]^T$. This subspace, parameterized by $\zb$, is an element of a Grassmanian manifold of dimension $p$. Now, let us rewrite $\Vb(\zb,\mathbb{I})$, using elementary operations, and with some abuse of notation, as
\begin{equation}
\Vb(\zb,\mathbb{I})=\begin{bmatrix} \Vb_1(\zb,\mathbb{I}) \\  \Vb_2(\zb,\mathbb{I})\\ \end{bmatrix},
\end{equation}
where $\Vb_1(\zb,\mathbb{I})\in\mathbb{C}^{p\times p}$ is invertible and $\Vb_2(\zb,\mathbb{I})\in\mathbb{C}^{(m-p)\times p}$. Then the basis matrix $\Ab(\zb,\mathbb{I})$ for the orthogonal subspace is the Hermitian transpose of
\begin{equation}\label{eq:ortho_general}
\Ab^H(\zb,\mathbb{I})=[-\Vb_2(\zb,\mathbb{I})\Vb_1^{-1}(\zb,\mathbb{I}) \ | \ \mathbf{I}_{m-p}].
\end{equation}
Although these $p$-dimensional characterizations of the signal and orthogonal subspaces have minimum parameterization $\zb\in\mathbb{C}^p$, it is not easy to solve the least squares problem (\ref{eq:min}) using these characterizations.

For an $m$-element uniform line array, a particular $p$-parameter characterization of $\Ab(\zb,\mathbb{I})$ exists that makes solving  \eqref{eq:min} relatively simple \cite{Prony1795}.  We will review this characterization in Section \ref{sec:review}. Then, we derive such suitable parameterizations of $\Ab(\zb,\mathbb{I})$ for two specific non-uniform arrays: sparse and co-prime.

%
\begin{itemize}
\item {\em Sparse array}: In this case, the location set $\mathbb{I}$ is given by $\mathbb{I}_s=\{0, d, 2d,  \dots, (m-2)d, M\}$, where $M$ and $d$ are co-prime integers, that is, $(M,d)=1$, and $d>1$. This array may be thought of as two subarrays. The first is a downsampled  version, by a factor $d$, of an $(m-1)d$-element uniform line array (ULA) with half wavelength interelement spacings. The second is a single sensor at location $M$ in the line array such that $M$ and $d$ are co-prime. We call this the sparse array because of the single element that sits apart from the origin of the first subarray. We note that $M$ need not be greater than $(m-2)d$.

\item {\em Co-prime array:} In this case, $\mathbb{I}=\mathbb{I}_1 \cup \mathbb{I}_2$, where $\mathbb{I}_1=\{0, m_2, 2m_2,\dots, (m_1-1)m_2\}$, $\mathbb{I}_2=\{m_1, 2m_1,\dots, (2m_2-1)m_1\}$, and $(m_1,m_2)=1$. Again the array is composed of two subarrays.  The first is an $m_1$-element ULA with interelement spacings of $m_2$ and sensor locations $\mathbb{I}_1$. The second is a $(2m_2-1)$-element ULA with interelement spacings of $m_1$ and sensor locations $\mathbb{I}_2$. This co-prime geometry was recently introduced in \cite{coprime1} and \cite{coprime2}.
\end{itemize}

\remark{2} In both cases, the co-prime constraint guarantees that aliasing ambiguities due to undersampling can be resolved. Although a sparse array can be viewed as a special case of a co-prime array, we consider them separately, because it is easier to first derive a suitable characterization of the orthogonal subspace $\langle \Ab \rangle$ for a sparse array, and then generalize it to a co-prime array. Our parameterizations are not minimal. They involve $2p$ parameters, instead of $p$, but as we will show in Section \ref{sec:main}, they are specifically designed to utilize modern methods of linear prediction and approximate least squares, such as IQML.


\remark{3} By now it should be clear that $\Ab(\zb,\mathbb{I})$, and therefore its parameterization, depend on both the mode vector $\zb$ and the array geometry $\mathbb{I}$. Therefore, from here on, we may drop $(\zb,\mathbb{I})$
and simply use $\Ab$, $\Vb$, $\langle \Ab \rangle$, and $\langle \Vb \rangle$.

\section{Characterization of the Orthogonal Subspace for Uniform Line Arrays}\label{sec:review}

Consider a uniform line array of $m$ equidistant sensors located at $\mathbb{I}_u=\{0,1,2,\dots,m-1\}$, taking measurements from the superposition of $p$ modes as in (\ref{eq:Main1}). The signal subspace in this case is characterized by the Vandermonde matrix $\Vb$ in (\ref{eq:Main2}) with $\mathbb{I}=\mathbb{I}_u$. To characterize the orthogonal subspace $\langle\Ab\rangle$, consider the polynomial $A(z)$:
\begin{align}\label{eq:AZ_unif}
A(z)&=\prod_{k=1}^{p}(1-z_kz^{-1})\nonumber \\
&=\sum_{i=0}^{p}a_iz^{-i};\hspace{5mm} a_0=1
\end{align}
which has  $(z_1,z_2,\dots, z_p)$ as its $p$ complex roots. The $(m-p)$ dimensional orthogonal subspace $\langle\Ab\rangle$ is spanned by the $m-p$ linearly independent columns of  $\Ab$:
\begin{equation}\label{eq:ortho}
\mathbf{A}= \begin{bmatrix}
  a_p & a_{p-1} & \cdots & a_1 & 1 & 0 &\cdots & 0 \\
   0&a_p &  & & &\ddots & \ddots &\vdots  \\
   \vdots&\ddots & \ddots & &  & & \ddots &\vdots   \\
   0 & \cdots & 0 & a_p & \cdots &\cdots & a_1 &1  \\
 \end{bmatrix}^H.
\end{equation}
Since $\Ab^H\Vb=\mathbf{0}$, and the columns of $\Ab$ are linearly independent, $\Vb$ and $\Ab$ span orthogonal subspaces $\langle\Vb\rangle$ and $\langle\Ab\rangle$ in $\mathbb{C}^m$. The above parameterization is at the heart of methods of linear prediction, approximate least squares, and IQML (see, e.g., \cite{Kumaresan86}--\nocite{Bresler86}\nocite{Kumaresan83}\cite{Kumaresan84}).

Using the $p$-parameter representation for $\langle \Ab\rangle$ in (\ref{eq:ortho}) we may re-write the least squares problem of (\ref{eq:min}) as
\begin{equation}\label{eq:min_par}
\hat{\mathbf{a}}=\mathop{\mathrm{argmin}}_{\mathbf{a}=[a_1,\dots,a_p]^T\in \mathbb{C}^{p}}\sum_{n=\pp{0}}^{\pp{N-1}}\yb^H[n]\mathbf{P}_{\Ab}\yb[n].
\end{equation}
There are many algorithms to approximately solve the nonlinear least squares problem in (\ref{eq:min_par}). One approach is to ignore the $(\Ab^H\Ab)^{-1}$ term in the projection matrix $\mathbf{P}_{\Ab}=\Ab(\Ab^H\Ab)^{-1}\Ab^H$ and solve the following modified least squares or linear prediction problem:
\begin{align}\label{eq:min_par2}
\hat{\mathbf{a}}=\mathop{\mathrm{argmin}}_{\mathbf{a}\in \mathbb{C}^{p}}\sum_{n=\pp{0}}^{\pp{N-1}}\yb^H[n]\Ab\Ab^H\yb[n].
\end{align}
The iterative quadratic maximum likelihood (IQML) algorithm (see \cite{Kumaresan86}, \cite{Bresler86}, and \cite{McClellan91}) is another method to approximately solve (\ref{eq:min_par}). In the $l$th iteration of IQML, the parameters $\mathbf{a}_l$ are estimated by iteratively minimizing the quadratic form
\begin{align}\label{eq:minpar3}
\hat{\mathbf{a}}_l=\mathop{\mathrm{argmin}}_{\mathbf{a}_l\in \mathbb{C}^{p}}~\mathbf{a}_l^H\left[\sum_{n=0}^{N-1}\Yb^H[n](\Ab_{l-1}^H\Ab_{l-1})^{-1}\Yb[n]\right]\mathbf{a}_l,
\end{align}
where $\Ab_{l-1}$ is formed as in (\ref{eq:ortho}) using the estimated parameters $\hat{\mathbf{a}}_{l-1}$ from iteration $(l-1)$ and $\yb[n]$ is the following $(m-p)\times p$ Hankel data matrix for snapshot $n$:
\begin{equation}
\Yb[n]= \begin{bmatrix}
  y_0[n] & \cdots & y_{p-1}[n] & y_{p}[n] \\
   y_1[n] & \cdots & y_{p}[n] & y_{p+1}[n] \\
   \vdots &  & \vdots & \vdots \\
  y_{m-1-p}[n] & \cdots & y_{m-2}[n] & y_{m-1}[n]
 \end{bmatrix}.
\end{equation}
After a number of these iterations the sequence $\{\hat{\mathbf{a}}_{l}\}$ converges to an estimate $\hat{\mathbf{a}}=[\hat{a}_1,\ldots,\hat{a}_p]\pp{^T}$. The polynomial $\hat{A}(z)=\sum_{i=0}^{p}\hat{a}_iz^{-i}$ is then formed from this estimate and its roots are taken as the mode estimates $(\hat{z}_1, \hat{z}_2, \dots, \hat{z}_p)$.
\section{Characterization of the Orthogonal Subspaces for Sparse and Co-prime Arrays}\label{sec:main}

In this section, we present simple characterizations of the orthogonal subspace $\langle\Ab\rangle$ for the sparse and co-prime arrays discussed in Section \ref{sec:PS}. Based on our characterizations, we adopt IQML for approximate least squares estimation of complex exponential modes in such arrays.

\subsection{Sparse Array}\label{sec:sparse}

Consider the sparse array described in Section \ref{sec:PS}. The set of sensor locations for this array is $\mathbb{I}_s=\{0, d, 2d,  \dots, (m-2)d, M\}$. The generalized Vandermonde matrix $\Vb$ in this case is
\begin{equation}\label{eq:vzis}
\Vb(\zb,\mathbb{I}_s)= \begin{bmatrix}
  1 & 1 & \cdots & 1 \\
  z_1^d & z_2^d & \cdots & z_p^d \\
  \vdots  & \vdots  & \ddots & \vdots  \\
  z_1^{(m-2)d} & z_2^{(m-2)d} & \cdots & z_p^{(m-2)d}\\
 z_1^M & z_2^M & \cdots & z_p^M
 \end{bmatrix}.
\end{equation}
For $d>1$ it is clear that without the use of the last sensor at location $M$, we cannot unambiguously estimate the modes, because any two modes $z_k$ and $z_k e^{j2\pi q/d}, ~ q=1,2,\cdots,d-1$ produce the same measurement. This is the aliasing problem for subsampled arrays.

To characterize the $(m-p)$-dimensional orthogonal subspace $\langle \Ab \rangle$, determined by the modes $\{z_k\}_{k=1}^{p}$, we first form the polynomial $A(z)$ from the $d$th powers of $z_k$, namely the $w_k=z_k^d$, $k=1,2,\dots,p$:
\begin{align}\label{eq:AZ}
A(z)&=\prod_{k=1}^{p}(1-w_kz^{-1})=\sum_{i=0}^{p}a_iz^{-i};\quad a_0=1.
\end{align}
Since $\{w_k\}_{k=1}^{p}$ are the roots of $A(z)$, the first $m-p-1$ columns $\mathbf{A}_0$ of $\mathbf{A}\in\mathbb{C}^{m\times (m-p)}$, which is to satisfy $\mathbf{A}^H\Vb=\mathbf{0}$, can be written as
\begin{equation}\label{eq:row1}
\mathbf{A}_0^H= \begin{bmatrix}
  a_p & a_{p-1} & \cdots & a_1 & 1 & 0 &\cdots & &0 \\
   0&a_p &  & & &\ddots & \ddots & &\vdots  \\
   \vdots&\ddots & \ddots & &  & & \ddots & &\vdots   \\
   0 & \cdots & 0 & a_p & \cdots & \cdots & a_1 & 1&0  \\
 \end{bmatrix}.
\end{equation}
But of course any mode of the form $z_{k}e^{j 2\pi q/d}$, $q=1,\ldots, d-1$, would produce the same $w_k$ and therefore the same $\Ab_0$. This is the ambiguity caused by aliasing.

Now, consider the polynomial
\begin{equation}\label{eq:BZ}
B(z)=z^M+\sum\limits_{i=1}^p b_iz^{(p-i)d}.
\end{equation}
Suppose the coefficient vector $\mathbf{b}=[b_1,b_2,\cdots,b_p]^T$ is such that the actual modes $\{z_k\}_{k=1}^{p}$ are the roots of $B(z)$. That is, $B(z_k)=0~\makebox{for}~k=1,2,\dots,p$. Then, since $M$ and $d$ are co-prime, for $1\leq q \leq d-1$ and $1\leq k \leq p$ we have
\begin{align}
B(z_ke^{j2\pi q/d})&=z_k^Me^{j2\pi Mq/d}+\sum\limits_{i=1}^p b_iz_k^{(p-i)d}\nonumber\\
&=z_k^M(e^{j2\pi Mq/d}-1)\nonumber\\
&\neq 0~~\makebox{for}~ q=1,2,\dots,d-1.
\end{align}
Therefore, the only common roots of $B(z)$, and the $d$th roots of $\{w_k\}_{k=1}^p$, are $\{z_k\}_{k=1}^p$, which are the actual modes to be estimated. In this way, $B(z)$ resolves the ambiguities.

Now suppose $\{w_k\}_{k=1}^{p}$ are known (or estimated). Then from (\ref{eq:BZ}), $\mathbf{b}$ can be found by solving the linear system of equations
\begin{equation}\label{eq:row2}
\begin{pmatrix}
  b_p &  b_{p-1} &  \hdots & b_1
 \end{pmatrix}
 \begin{pmatrix}
  1 & 1 & \cdots & 1  \\
   z_1^d & z_2^d & \cdots &  z_p^d \\
  \vdots  & \vdots  & \ddots & \vdots  \\
  z_1^{(p-1)d} & z_2^{(p-1)d} & \cdots &  z_p^{(p-1)d}
 \end{pmatrix}
=
-\begin{pmatrix}
z_1^M &  z_2^M &  \hdots & z_p^M
 \end{pmatrix},
\end{equation}
which if $z_i^d\neq z_j^d$ for $i \neq j$ (as we assume) has a unique solution.

Using the $2p$ coefficients $\{a_i\}_{i=1}^{p}$ and $\{b_i\}_{i=1}^p$, we can characterize $\langle \Ab \rangle$ by writing $\Ab\in\mathbb{C}^{m  \times (m-p)}$ as
\begin{equation}\label{eq:char}
\mathbf{A}=\begin{bmatrix}
  a_p & a_{p-1} & \cdots & a_1 & 1 & 0 &\cdots & 0 \\
   0&a_p &  & & &\ddots & \ddots &\vdots  \\
   \vdots&\ddots & \ddots & &  & & \ddots &\vdots   \\
   0 & \cdots & 0 & a_p & \cdots & a_1 & 1 &0  \\
b_p& b_{p-1}&\cdots&b_1&0&\cdots&0&1
\end{bmatrix}^H.
\end{equation}

To estimate $\mathbf{a}=[a_1,\cdots,a_p]^T$ and $\mathbf{b}=[b_1,\cdots,b_p]^T$, we need to solve the following problem:
\begin{align}\label{eq:mincom}
\min_{\mathbf{a,b}}\sum_{n=0}^{N-1}\yb^H[n]\mathbf{P}_{\Ab}\yb[n],
\end{align}
We approximate the solution to this problem in two steps. First, we ignore the last column of $\Ab$ and estimate $\mathbf{a}$ as
\begin{align}\label{eq:min1}
\hat{\mathbf{a}}=\mathop{\mathrm{argmin}}_{\mathbf{a}}\sum_{n=0}^{N-1}\yb^H[n]\mathbf{P}_{\Ab_0}\yb[n].
\end{align}
In the noiseless case, where the $\yb[n]$, $n=0,1, \ldots, N-1$, lie in $\langle\Vb\rangle$, it can be shown that if $m\geq 2p+1$ then the solution to (\ref{eq:min1}) is unique and yields the coefficients of the polynomial $A(z)$ with roots $(w_1, w_2, \dots , w_p)$. See Appendix A.

The minimization problem in (\ref{eq:min1}) can be solved using IQML. Now, given $\hat{\mathbf{a}}$, we form the polynomial
\begin{align}
\hat{A}(z)&=1+\sum_{i=1}^{p}\hat{a}_iz^{-i}\nonumber\\
&=\prod_{k=1}^{p}(1-\hat{w}_kz^{-1})
\end{align}
and derive its roots as $\{\hat{w}_k\}_{k=1}^p$. But we know from the structure of the problem that $\hat{w}_k=\hat{z}_k^d$, and any of the $d$-th roots of $\hat{z}_k^d$ is a candidate solution. Therefore, we construct the candidate set,
\begin{align}\label{eq:R}
\mathcal{R}=\{&(\hat{z}_1e^{j2\pi q_1/d},\hat{z}_2e^{j2\pi q_2/d},\cdots,\hat{z}_pe^{j2\pi q_p/d}) | 0\leq q_1, q_2, \dots, q_p\leq d-1\},
\end{align}
which contains all modes and their aliased versions.

In the second step, to find the $p$ actual modes and resolve aliasing ambiguities, we solve the following constrained linear prediction problem:
\begin{align}\label{eq:minB}
\mathbf{\hat{b}}&=\mathop{\mathrm{argmin}}_{\boldsymbol{\zeta}}\sum_{n=0}^{N-1}|y_{m-1}[n]+ \boldsymbol{\zeta}^T\mathbf{u}[n]|^2\nonumber \\
& \mathrm{s.t.} \ \  B_{\boldsymbol{\zeta}}(\hat{\zb})=0, \ \ \hat{\zb}\in\mathcal{R},
\end{align}
where $\mathbf{u}[n]=[y_0[n], y_1[n], \dots, y_{p-1}[n]]^T$, and the polynomial $B_{\boldsymbol{\zeta}}(z)$ is obtained from replacing $\mathbf{b}$ by $\boldsymbol{\zeta}$ in (\ref{eq:BZ}).

In the noiseless case, where $\yb[n]$, $n=0,1, \ldots, N-1$ lie in $\langle\Vb\rangle$, the solution $\hat{\mathbf{b}}$ to (\ref{eq:minB}) satisfies (\ref{eq:row2}) and yields the actual modes. See Appendix B.

Our algorithm for estimating modes in a sparse array may be summarized in the following steps:
\begin{enumerate}
\item Estimate $\hat{\mathbf{a}}=[\hat{a}_1, \hat{a}_2, \dots, \hat{a}_p]^T$ from (\ref{eq:min1}) using IQML;

\item Root $\hat{A}(z)$ to return roots $\{\hat{w}_k\}_{k=1}^p$. Then, recognizing that \pp{the $d$th roots of $\hat{w}_k$ are $\hat{z}_ke^{j2\pi q/d}$ for some $q\in\{0,1,2,\dots,d-1\}$}, form the set of candidate modes $\mathcal{R}$ as in (\ref{eq:R});

\item Solve (\ref{eq:minB}) for $\hat{\mathbf{b}}$;

\item Intersect the roots of $\hat{B}(z)$ with $\mathcal{R}$.
\end{enumerate}

\subsection{Co-prime Array}\label{sec:coprime}

Consider an $m=m_1+2 m_2-1$ element co-prime array, consisting of two uniform subarrays: one with $m_1$ elements at locations $\mathbb{I}_1=\{0, m_2, 2m_2,\dots, (m_1-1)m_2\}$ and the other with $2m_2-1$ elements at locations $\mathbb{I}_2=\{m_1, 2m_1,\dots, (2m_2-1)m_1\}$, where $(m_1,m_2)=1$ and $m_1>m_2$. In this case, the generalized Vandermonde matrix $\Vb\in\mathbb{C}^{m\times p}$ of modes may be partitioned as
\begin{equation}
\Vb= \begin{bmatrix}\Vb(\zb,\mathbb{I}_1) \\ \Vb(\zb,\mathbb{I}_2) \end{bmatrix},
\end{equation}
where
\begin{equation}
\Vb(\zb,\mathbb{I}_1)= \begin{bmatrix}
  1 & 1 & \cdots & 1 \\
  z_1^{m_2} & z_2^{m_2} & \cdots & z_p^{m_2} \\
  z_1^{2m_2} & z_2^{2m_2} & \cdots & z_p^{2m_2} \\
  \vdots  & \vdots  & \ddots & \vdots  \\
  z_1^{(m_1-1)m_2} & z_2^{(m_1-1)m_2} & \cdots & z_p^{(m_1-1)m_2}\\
 \end{bmatrix}
\end{equation}
and
\begin{equation}
\Vb(\zb,\mathbb{I}_2)= \begin{bmatrix}
  z_1^{m_1} & z_2^{m_1} & \cdots & z_p^{m_1} \\
  z_1^{2m_1} & z_2^{2m_1} & \cdots & z_p^{2m_1} \\
  \vdots  & \vdots  & \ddots & \vdots  \\
  z_1^{(2m_2-1)m_1} & z_2^{(2m_2-1)m_1} & \cdots & z_p^{(2m_2-1)m_1}\\
 \end{bmatrix}
\end{equation}
are the Vandermonde matrices for the two individual subarrays of the co-prime array.

Let $\Ab_1\in\mathbb{C}^{m_1\times(m_1-p)}$ and $\mathbf{B}_1\in\mathbb{C}^{(2m_2-1)\times (2m_2-1-p)}$ be matrices that are orthogonal to $\Vb(\zb,\mathbb{I}_1)$ and $\Vb(\zb,\mathbb{I}_2)$, respectively. That is, $\Ab_1^H\Vb(\zb,\mathbb{I}_1)=\mathbf{0}$ and $\mathbf{B}_1^H\Vb(\zb,\mathbb{I}_2)=\mathbf{0}$. Following our results in the sparse case, we may parameterize $\Ab_1\in\mathbb{C}^{m_1\times(m_1-p)}$ as
\begin{equation}\label{eq:row1_coprime}
\mathbf{A}_1^H= \begin{bmatrix}
  a_p & a_{p-1} & \cdots & a_1 & 1 & \cdots &0 \\
   0&a_p &\cdots& & &\cdots & 0 \\
   \vdots&\ddots & \ddots & & \cdots & & \vdots\\
   0 & \cdots & 0 & a_p & \cdots & a_1 & 1\\
 \end{bmatrix},
\end{equation}
where $\{a_i\}_{i=1}^{p}$ are the coefficients of a polynomial $A(z)$, whose roots are $w_k=z_k^{m_2}$, $k=1,2,\dots,p$. That is,
\begin{align}\label{eq:AZ_coprime}
A(z)&=\prod_{k=1}^{p}(1-w_kz^{-1})\nonumber \\
&=\sum_{i=0}^{p}a_iz^{-i},\ \  a_0=1.
\end{align}
Similarly, we parameterize $\Bb_1\in\mathbb{C}^{(2m_2-1)\times (2m_2-1-p)}$ as
\begin{equation}\label{eq:row2_coprime}
\mathbf{B}_1^H= \begin{bmatrix}
  b_p & b_{p-1} & \cdots & b_1 & 1 & \cdots &0 \\
   0&b_p &\cdots& & &\cdots & 0 \\
   \vdots&\ddots & \ddots & & \cdots & & \vdots\\
   0 & \cdots & 0 & b_p & \cdots & b_1 & 1\\
 \end{bmatrix},
\end{equation}
where $\{b_i\}_{i=1}^{p}$ are the coefficients of a polynomial $B(z)$, whose roots are $s_k=z_k^{m_1}$, $k=1,2,\dots,p$. That is,
\begin{align}\label{eq:BZ_coprime}
B(z)&=\prod_{k=1}^{p}(1-s_kz^{-1})\nonumber \\
&=\sum_{i=0}^{p}b_iz^{-i}, \ \ b_0=1.
\end{align}

Note that we still need $p$ more independent columns to fully characterize the basis matrix $\Ab$ for the orthogonal subspace $\langle \Ab \rangle$. However, using our partial characterization, we can estimate the modes (with no aliasing ambiguities) in the following steps:

\begin{enumerate}
\item Separate the measurements of the two subarrays as $\mathbf{u}[n]=\{\yb_i[n]|i\in\mathbb{I}_1\}$ and $\mathbf{v}[n]=\{\yb_i[n]|i\in\mathbb{I}_2\}$;
\item Estimate $\hat{\mathbf{a}}=[\hat{a}_1, \hat{a}_2, \dots, \hat{a}_p]^T$ using IQML on $\mathbf{u}[n]$;
\item Root $\hat{A}(z)$ to return the roots $\{\hat{w}_k\}_{k=1}^p$. Then, recognizing that \pp{the $m_2$th roots of $\hat{w}_k$ are $\hat{z}_ke^{j2\pi q/m_2}$ for some $q\in\{0,1,\dots,m_2-1\}$}, form the set of candidate modes $\mathcal{R}_1$ as
\begin{equation}\label{eq:R1}
\mathcal{R}_1=\{(\hat{z}_1e^{j2\pi k_1/m_2},\hat{z}_2e^{j2\pi k_2/m_2},\cdots,\hat{z}_pe^{j2\pi k_p/m_2}) \hspace{2mm}| \hspace{2mm} 0\leq k_1, k_2, \dots, k_p\leq m_2-1\};
\end{equation}
\item Estimate $\hat{\mathbf{b}}=[\hat{b}_1, \hat{b}_2, \dots, \hat{b}_p]^T$ using IQML on $\mathbf{v}[n]$;
\item Root $\hat{B}(z)$ to return the roots $\{\hat{s}_k\}_{k=1}^p$. Then, recognizing that \pp{the $m_1$th roots of $\hat{s}_k$ are $\hat{z}_ke^{j2\pi q/m_1}$ for some $q\in\{0,1,\dots,m_1-1\}$}, form the set of candidate modes $\mathcal{R}_2$ as
\begin{equation}\label{eq:R1}
\mathcal{R}_2=\{(\hat{z}_1e^{j2\pi k_1/m_1},\hat{z}_2e^{j2\pi k_2/m_1},\cdots,\hat{z}_pe^{j2\pi k_p/m_1}) \hspace{2mm}| \hspace{2mm} 0\leq k_1, k_2, \dots, k_p\leq m_1-1\};
\end{equation}
\item Intersect $\mathcal{R}_1$ and $\mathcal{R}_2$, in other words look for the closest (based on the Euclidean metric) $p$ members of the set $\mathcal{R}_1$ to the set $\mathcal{R}_2$.
\end{enumerate}

\remark{4} To complete the $2p-$parameterization of the basis matrix $\Ab$ for the orthogonal subspace $\langle \Ab \rangle$, consider the standard representation of $\Ab$ given in (\ref{eq:ortho_general}). Define $\Vb_1=\Vb(\zb,\mathbb{I}_{p_1})$ and $\Vb_2=\Vb(\zb,\mathbb{I}_{p_2})$ where $\mathbb{I}_{p_1}=\{0,m_2,\dots,(p-1)m_2\}$ and $\mathbb{I}_{p_2}=\{m_1,2m_1,\dots,pm_1\}$. Then, from (\ref{eq:ortho_general}) the $p$ remaining columns of $\Ab$ may be represented in $\mathbf{C}_1\in\mathbb{C}^{n\times p}$ as:
\begin{equation}\label{eq:C0}
\Cb_1^H=[\mathbf{C}_0^H|\hspace{2mm} \mathbf{0}_{p\times(m_1-p)}\hspace{1mm}|\hspace{2mm} \mathbf{I}_p\hspace{1mm}| \hspace{2mm} \mathbf{0}_{p\times(2m_2-1-p)}]
\end{equation}
where $\mathbf{0}_{k\times l}$ denotes a $k\times l$ matrix with zero entries, $\mathbf{I}_p$ is the $p\times p$ identity matrix, and
\begin{equation}\label{eq:C1}
\Cb_0^H=-\Vb(\zb,\mathbb{I}_{p_2})\Vb^{-1}(\zb,\mathbb{I}_{p_1})\in\mathbb{C}^{p\times p}.
\end{equation}
From (\ref{eq:C0}) and (\ref{eq:C1}) we can see that $\mathbf{C}_1$ only depends on $\{z_k^{m_1}\}_{k=1}^p$ and $\{z_k^{m_2}\}_{k=1}^p$ which are obtained from $\mathbf{a}$ and $\mathbf{b}$ by rooting $A_{\mathbf{a}}(z)$ and $B_{\mathbf{b}}(z)$ in (\ref{eq:AZ_coprime}) and (\ref{eq:BZ_coprime}), respectively. Therefore, the full, and minimally parameterized, characterization of the orthogonal subspace for the co-prime array may be written as

\begin{equation}
\Ab=\left[
\begin{array}{cc|c}

  \Ab_1 & 0&\vspace{-3mm}\\
   & & \Cb_1\vspace{-3mm}\\
0 & \Bb_1 &

\end{array}
\right].
\end{equation}
We note that we do not need this full characterization for estimating the modes. The partial characterization using $\Ab_1$ and $\Bb_1$ suffices, at the expense of $p$ fitting equations.

\section{Numerical Results}

In this section we present numerical results for the estimation of damped complex exponential modes in co-prime, sparse and uniform line arrays. We consider a ULA of $50$ elements. We form our co-prime and sparse arrays with $14$ elements by subsampling this ULA. For the sparse array, we subsample the measurements of the ULA by a factor of $d=4$ and place a sensor at $M=3$. For the co-prime array, the first subarray includes $m_1=7$ elements with interelement spacing of $m_2=4$, and the second subarray includes $2m_2-1=7$ elements with interelement spacing of $m_1=7$.

\begin{figure}
\centering
\includegraphics[width=115mm]{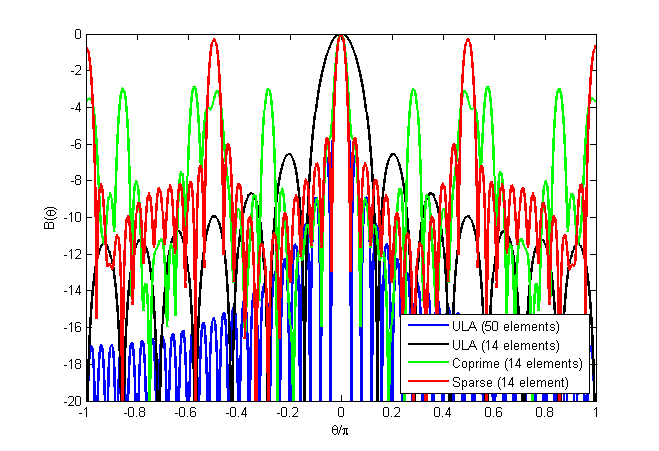}
\caption{Beampatterns for ULAs with $14$ and $50$ elements, a sparse array with $14$ elements, $d=4$, and $M=3$, and a co-prime array with $14$ elements, $m_1=7$, and $m_2=4$. }\label{fig:Beampattern}
\end{figure}

It is insightful to first look at the beampatterns of sparse, co-prime, and uniform line arrays for the problem of estimating undamped modes. In this case, the beam pattern $B(\theta)$ is
\begin{equation}
B(\theta)=\sum_{l=0}^{m-1}e^{ji_l\theta}.
\end{equation}
Figure \ref{fig:Beampattern} shows the beam patterns for different array geometries. Although the co-prime and sparse arrays of $14$ elements have the same aperture and the same main lobe width as the ULA with $50$ elements, we see that they suffer from higher sidelobes, suggesting that there will be performance losses in resolving closely spaced modes using these arrays, relative to the ULA.


Let us also look at numerical results for the Cram\'{e}r-Rao bound (CRB) associated with the co-prime, sparse and uniform line arrays (See Appendix C). Figure \ref{fig:CRB_3D} shows the CRB in the estimation of the mode $z_1=1$ in the presence of an interfering mode $z_2$. The per sensor SNR is $10$ dB. As the interfering mode $z_2$ gets closer to $z_1$, the CRB in the estimation of the $z_1$ increases. The CRB for the sparse and co-prime arrays are similar, but they are higher than the CRB for the ULA. Since the aperture of the three arrays are equal, the fewer number of sensors in the sparse and co-prime arrays can be considered  the only reason for this difference in the CRB.
\begin{figure}
\begin{center}
\begin{tabular}{c}
\includegraphics[width=3.5in]{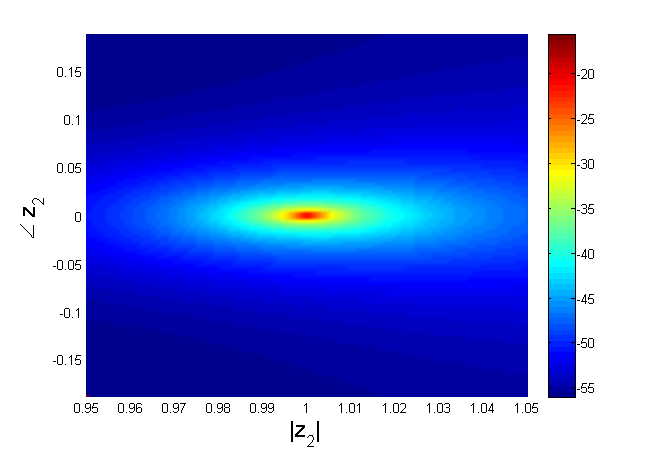}\\
(a) ULA\\
\includegraphics[width=3.5in]{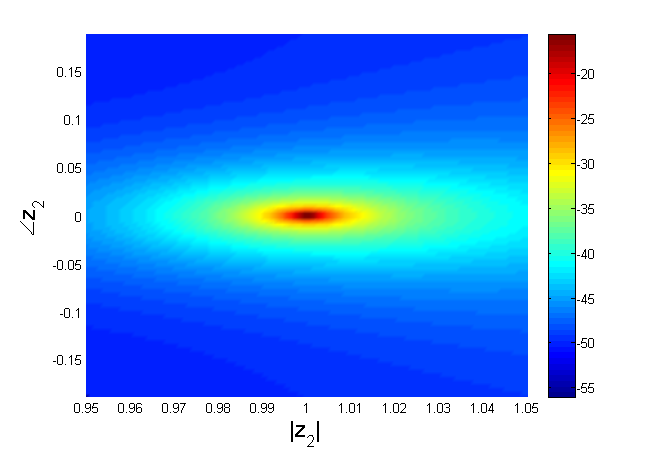}\\
(b) Sparse array\\
\includegraphics[width=3.5in]{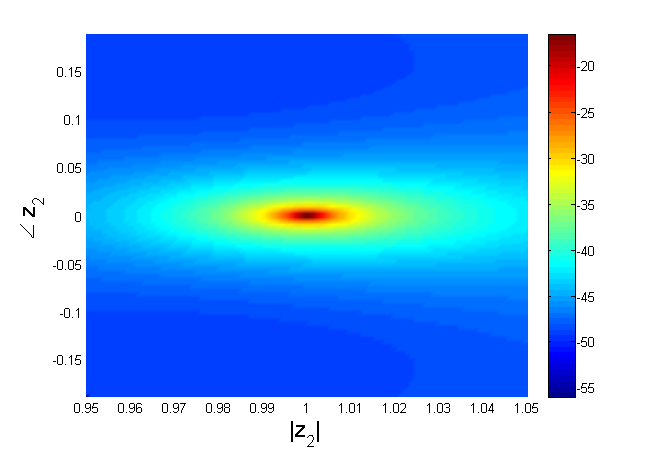}\\
 (c) Co-prime array
\end{tabular}
\end{center}
\caption{The CRB in dB for estimating $z_1=1$ in the presence of an interfering mode $z_2$: (a) ULA with $50$ elements. (b) sparse array with $14$ elements $d=4$ and $M=3$. (c) co-prime array with $23$ elements $m_1=7$ and $m_2=4$. For all arrays per sensor SNR is $10$ dB.}\label{fig:CRB_3D}
\end{figure}

\begin{figure}
\begin{tabular}{cc}
  \hspace{-0.7cm}\includegraphics[width=85mm]{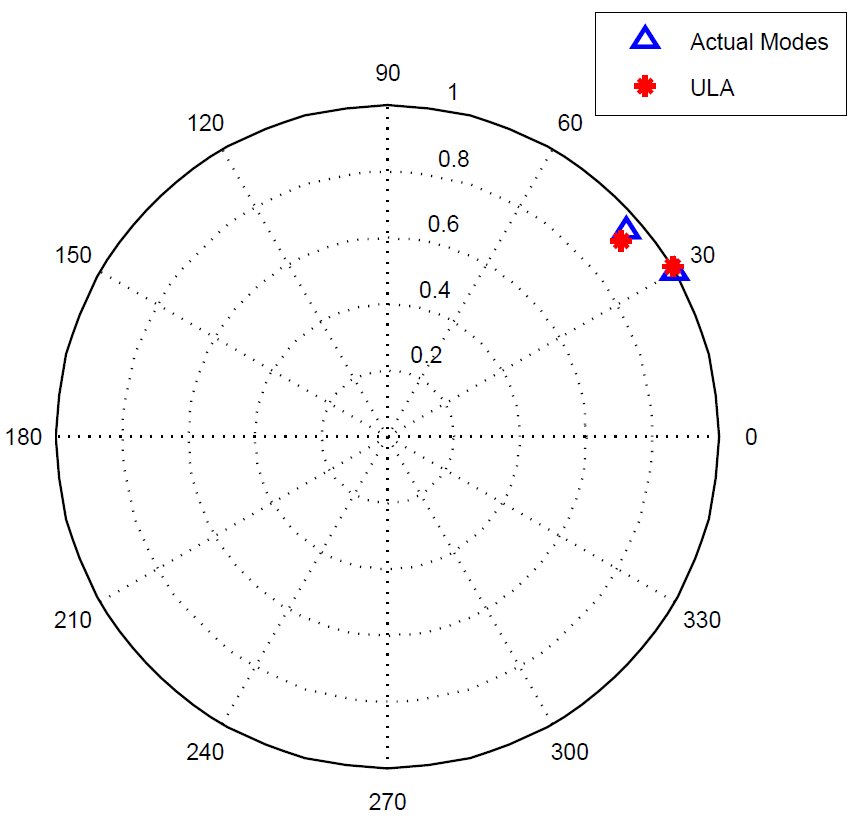} &   \hspace{-0.48cm}\includegraphics[width=85mm]{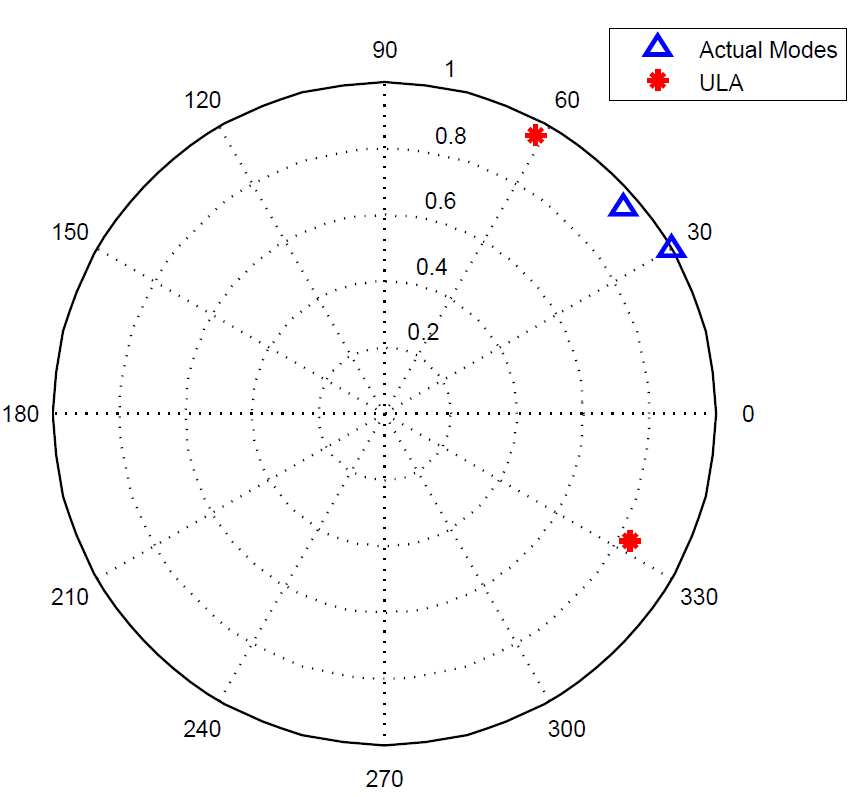} \\
\hspace{-1.4cm}(a)  & \hspace{-1.3cm}(b)
\end{tabular}
\caption{Estimating two closely spaced modes $z_1=e^{j0.52}$ and $z_2=0.95e^{j0.69}$ using a ULA with $50$ elements: (a) Per sensor SNR $=0$ dB. (b) Per sensor SNR $=-5$ dB.}\label{fig:LMSNR_ULA}
\end{figure}

\begin{figure}
\begin{tabular}{cc}
  \hspace{-0.7cm}\includegraphics[width=85mm]{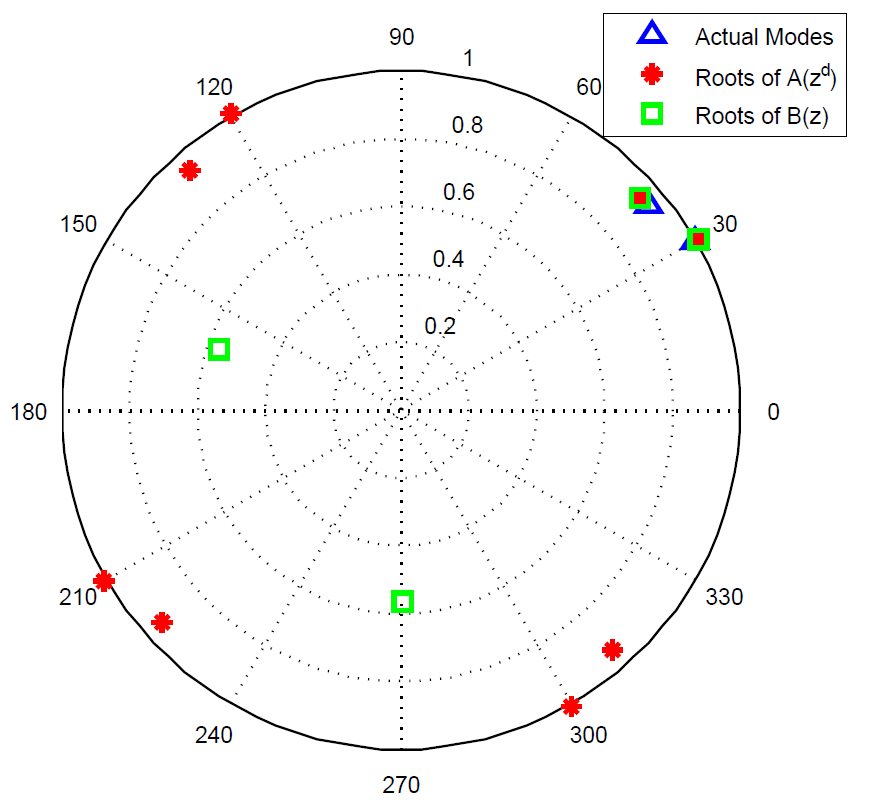} &   \hspace{-0.7cm}\includegraphics[width=85mm]{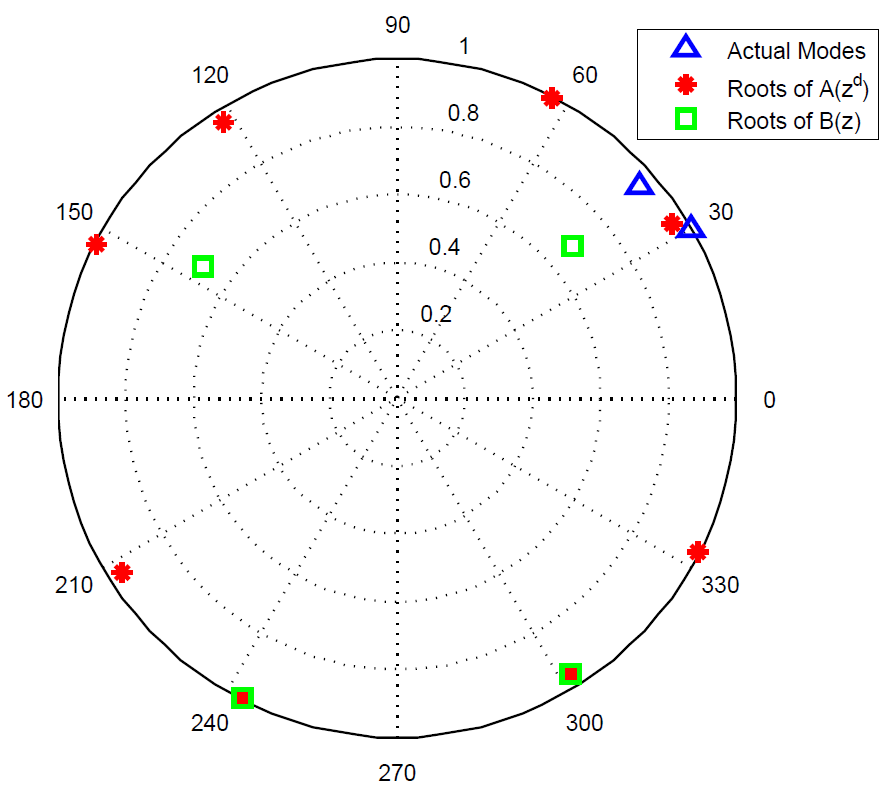} \\
\hspace{-1.4cm}(a)  & \hspace{-1.4cm}(b)
\end{tabular}
\caption{Estimating two closely spaced modes $z_1=e^{j0.52}$ and $z_2=0.95e^{j0.69}$ using a sparse array with $14$ elements, $d=4$ and $M=3$: (a) Per sensor SNR $=5$ dB (b) Per sensor SNR $=0$ dB.}\label{fig:LMSNR_Sparse}
\end{figure}

\begin{figure}
\begin{tabular}{cc}
  \hspace{-0.7cm}\includegraphics[width=88mm]{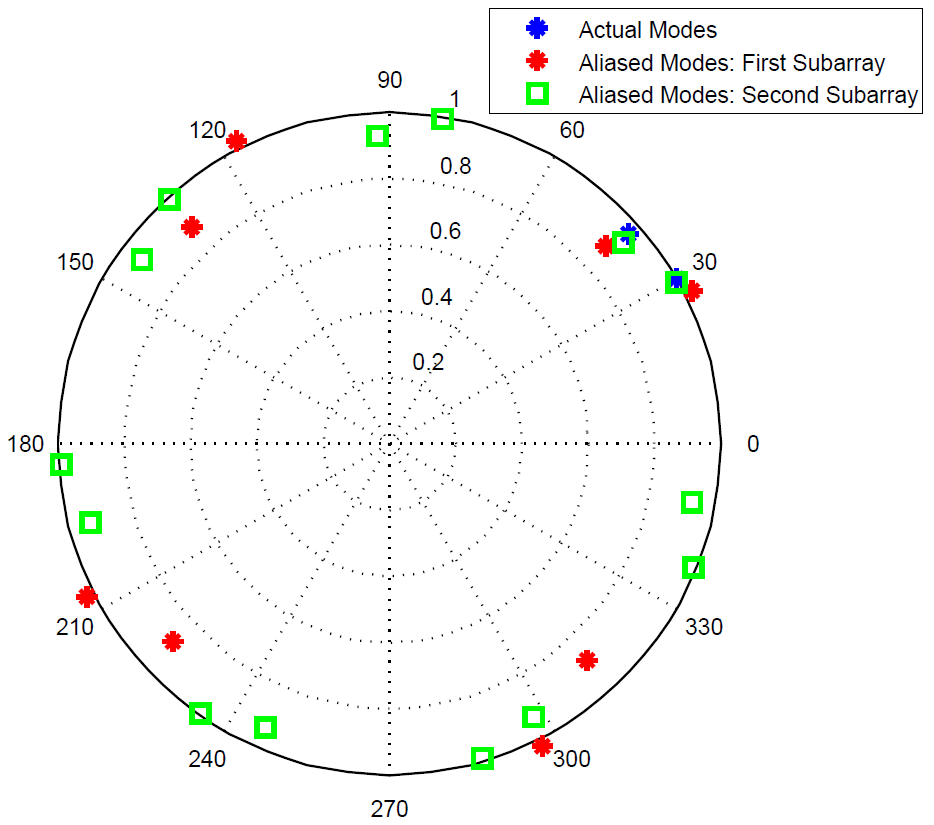} &   \hspace{-0.5cm}\includegraphics[width=88mm]{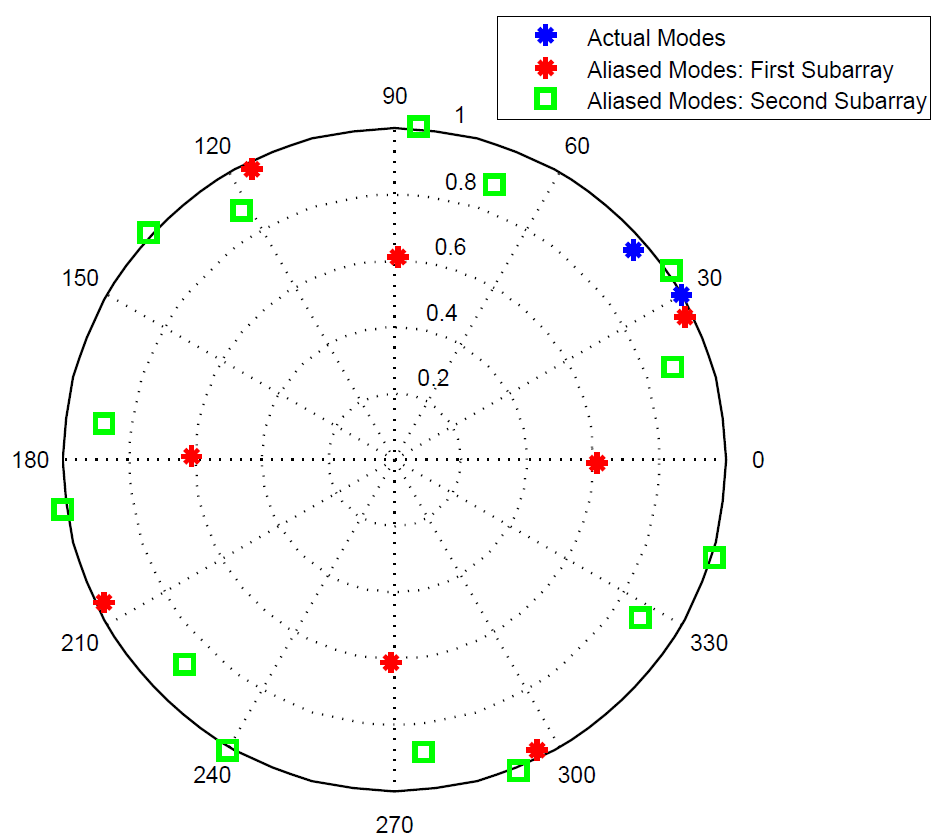} \\
\hspace{-2cm}(a)  & \hspace{-1.8cm}(b)
\end{tabular}
\caption{Estimating two closely spaced modes $z_1=e^{j0.52}$ and $z_2=0.95e^{j0.69}$ using a co-prime array with $14$ elements, $m_1=7$ and $m_2=4$: (a) Per sensor SNR $=5$ dB (b) Per sensor SNR $=0$ dB.}\label{fig:LMSNR_Cop}
\end{figure}

We now consider the performance of the approximate least squares estimation methods, shown in Figs.~\ref{fig:LMSNR_ULA}-\ref{fig:LMSNR_Cop}. The two modes to be estimated here are $z_1=e^{j0.52}$ and $z_2=0.95e^{j0.69}$. We choose the per sensor SNR values for the sparse and co-prime arrays to be $5$ dB higher than the SNR for the ULA, based on our insight in \cite{Pakrooh13b} about the threshold SNR for ULA and co-prime arrays. When SNR is decreased, there comes a point where some of the components of the orthogonal subspace better approximate the measurements than some of the components of the signal subspace. This leads to a performance breakdown in the estimation of the modes. The SNR at which this catastrophic breakdown occurs is called the threshold SNR (see \cite{Scharf95} and \cite{Pakrooh13b}). For the compression ratio of $50/14$, the threshold SNR for the co-prime and sparse arrays is almost $5$ dB more than its value for the ULA, which is a consequence of the subsamplings by these arrays. We emphasize that any compression increases the SNR threshold. The use of a co-prime or sparse array instead of a dense uniform line array is only justified in applications where SNR is high enough for the desired estimation resolution, or when SNR can be built up from temporal snapshots using long observation periods. The latter of course requires the scene to remain stationary over the longer estimation period.

\section*{Appendix A}

Let $m>2p$ and $z_i^d\neq z_j^d$ for $i\neq j$. Also, let $\mathbf{a}=(a_1, a_2, \dots, a_p)^T$ be the solution to (\ref{eq:min1}). That is,
\begin{align}\label{eq:min1_apendix}
\mathbf{a}=\mathop{\mathrm{argmin}}_{\boldsymbol{\alpha}}\sum_{n=0}^{N-1}\yb^H[n]\mathbf{P}_{\Ab_0(\boldsymbol{\alpha})}
\yb[n],
\end{align}
where ${\Ab_0}(\boldsymbol{\alpha})$ denotes an $\Ab_0$ of the form (\ref{eq:row1}), with $\alpha_i$'s replacing $a_i$'s for $i=1,2,\ldots,p$.
We wish to show that in the noiseless case the roots of the polynomial $A(z)=1+\sum_{i=1}^{p}a_iz^{-i}$  are $w_k=z_k^d$ for $k=1,2,\dots,p$ (the $d$th power of the actual modes).

Without loss of generality we assume that $N=1$. Let $\yb=\yb[0]$. In the noiseless case, the minimum value of the objective function $\yb^H\mathbf{P}_{\Ab_0(\boldsymbol{\alpha})}\yb$ is zero, and because $\Ab_0^H(\boldsymbol{\alpha})\Ab_0(\boldsymbol{\alpha})$ is always full rank, for the solution vector $\mathbf{a}$, we have $\Ab_0^H(\mathbf{a})\yb=\mathbf{0}$. Now, in the noiseless case, $\yb=\Vb(\zb,\mathbb{I}_s)\xb$, where $\Vb(\zb,\mathbb{I}_s)$ is given by \eqref{eq:vzis} and $\xb=[x_1,x_2,\ldots,x_p]^T$ is the vector of mode weights. Therefore, we have
\begin{equation}
\Ab_0^H(\mathbf{a})\Vb(\zb,\mathbb{I}_s)\xb=\mathbf{0},
\end{equation}
which we can reorder to get
\begin{equation}\label{eq:App1}
\begin{bmatrix}
  1 & 1 & \cdots &  1 \\
   z_1^d & z_2^d & \cdots &  z_p^d \\
  \vdots  & \vdots  & \ddots & \vdots  \\
  z_1^{(m-p-1)d} & z_2^{(m-p-1)d} & \cdots &  z_p^{(m-p-1)d}
 \end{bmatrix}
\begin{bmatrix}
x_1 & 0 & \cdots & 0\\
0 & x_2 & \cdots & 0\\
\vdots & \vdots & \ddots & \vdots\\
0 & 0 & \cdots & x_p
\end{bmatrix}
\begin{bmatrix}
A(w_1) \\
  A(w_2) \\
  \vdots\\
A(w_p)
 \end{bmatrix}
=
\mathbf{0}.
\end{equation}
Because the matrix on the left hand side of (\ref{eq:App1}) is a full column rank Vandermonde matrix, and the diagonal matrix in the middle is nonsingular (by the assumption of having $p$ actual modes), the above equality holds iff
\begin{equation}
A(w_k)= 0 \hspace{3mm} \textrm{for}  \hspace{2mm}  k=1,\dots,p.
\end{equation}


\section*{Appendix B}

Let $m>2p$ and $z_i^d\neq z_j^d$ for $i\neq j$. Also, let $\boldsymbol{\beta}$ be the solution to (\ref{eq:minB}) in the noiseless case. Here we use $\boldsymbol{\beta}$ instead of $\hat{\mathbf{b}}$ for the solution to distinguish noiseless and noisy cases. We show $\boldsymbol{\beta}$ solves (\ref{eq:row2}) and therefore resolves aliasing.

Again, without loss of generality, we assume $N=1$. Based on our argument in Appendix A, in the noiseless case we have
\begin{equation}
\mathcal{R}=\{(z_1e^{j2\pi k_1/d},z_2e^{j2\pi k_2/d},\cdots,z_pe^{j2\pi k_p/d}) \hspace{2mm}| \hspace{2mm} 0\leq k_1, k_2, \dots, k_p\leq d-1\},
\end{equation}
where $\{z_i\}_{i=1}^{p}$ are the actual modes. In this case (\ref{eq:minB}) can be rewritten as:
\begin{equation}\label{eq:minD}
\boldsymbol{\beta}=\arg\min_{\boldsymbol{\zeta}}|y_{m-1}+ \boldsymbol{\zeta}^T\mathbf{u}|^2\hspace{5mm} \mathrm{s.t.} \hspace{2mm} \Vb_p^T(\boldsymbol{\eta})\boldsymbol{\zeta}=-\boldsymbol{\eta}^{\odot M}, \hspace{2mm} \boldsymbol{\eta}\in\mathcal{R},
\end{equation}
where $\boldsymbol{\eta}=(\eta_1, \eta_2, \ldots, \eta_p)$, $\boldsymbol{\eta}^{\odot M}=[\eta_1^M,\eta_2^M,\ldots,\eta_p^M]^T$ and
\begin{equation}
\Vb_p(\boldsymbol{\eta})=\begin{bmatrix}
  1 & 1 & \cdots &  1 \\
   \eta_1^d & \eta_2^d & \cdots & \eta_p^d \\
  \vdots  & \vdots  & \ddots & \vdots  \\
  \eta_1^{(p-1)d} & \eta_2^{(p-1)d} & \cdots &  \eta_p^{(p-1)d}
 \end{bmatrix}.
\end{equation}
In the noiseless case, $\mathbf{u}=\Vb_p(\zb)\xb$, $y_{m-1}=\sum_{i=1}^{p}x_i z_i^M$ and $\Vb_p(\boldsymbol{\eta})=\Vb_p(\zb)$. Therefore, we have
\begin{align}\label{eq:Eb}
|y_{m-1}+ \boldsymbol{\beta}^T\mathbf{u}|^2 &
=|\sum_{i=1}^{p}x_iz_i^M+\boldsymbol{\beta}^T\Vb_p(\zb)\xb|\nonumber\\
&=|\sum_{i=1}^{p}x_i(z_i^M-\eta_i^M)|\geq 0.
\end{align}
Now, because $\boldsymbol{\beta}$ is the solution to (\ref{eq:minD}), then in the noiseless case $|y_{m-1}+ \boldsymbol{\beta}^T\mathbf{u}|^2=0$ and from (\ref{eq:Eb}) we have $\boldsymbol{\eta}^{\odot M}=\mathbf{z}^{\odot M}$ almost surely. Therefore,
\begin{align}
\boldsymbol{\beta}&=-(\Vb_P^{T}(\boldsymbol{\eta}))^{-1}\boldsymbol{\eta}^{\odot M}\nonumber\\
&=-(\Vb_P^{T}(\zb))^{-1}\zb^{\odot M} \nonumber\\
&=\mathbf{b},
\end{align}
where $\mathbf{b}$ is the solution to (\ref{eq:row2}). $~\blacksquare$


\section*{Appendix C}

The Fisher information matrix for the measurement model in (\ref{eq:Main1}) may be written as
\begin{equation}
\mathbf{J}(\zb)=\sum_{n=1}^{N} \mathbf{J}_n(\zb), \quad n=0,1,\ldots, N-1,
\end{equation}
where $\mathbf{J}_n(\zb)$ is the Fisher information matrix for the estimation of the modes $\mathbf{z}=[z_1, z_2, \cdots, z_p]^T$ from $\yb[n]$ in (\ref{eq:Main1}). That is,
\begin{equation}
\mathbf{J}_n(\zb)=\frac{1}{\sigma^2}\Gb_n^H(\zb)\Gb_n(\zb),
\end{equation}
where $\Gb_n(\zb)=[\mathbf{g}_1[n], \mathbf{g}_2[n], \cdots ,\mathbf{g}_p[n]]$, and
\begin{equation}
\mathbf{g}_l[n]=\begin{bmatrix}
  i_0z_l^{i_0-1}  \\
   i_1z_l^{i_1-1} \\
   \vdots \\
  i_{(m-1)}z_l^{i_{(m-1)}-1}
\end{bmatrix}x_l[n]
\end{equation}
is the $l$th sensitivity vector, \pp{for $1\leq l\leq p$}. The CRB for the estimation of the $k$th \pp{mode $z_k$} is the $k$th diagonal element of $\mathbf{J}^{-1}(\zb)$.

\section{Acknowledgment}
The authors thank Prof. Chris Peterson for pointing out the minimal, $p$-dimensional, characterization of the orthogonal subspace in (\ref{eq:ortho_general}).

\bibliographystyle{IEEEtran}
\bibliography{ref_Journal}

\end{document}